\documentclass[a4paper,runninghead]{llncs}

\usepackage{graphicx}
\usepackage{latexsym,amssymb,amsmath,amsfonts}
\usepackage{url}
\usepackage{float}
\usepackage{multicols}
\usepackage{alltt}
\usepackage{enumerate}
\usepackage{makeidx}

\newcommand{\benum}{\begin{enumerate}}
\newcommand{\eenum}{\end{enumerate}}
\newcommand{\nc}{\newcommand}
\newcommand{\rnc}{\renewcommand}
\nc{\teta}{\tilde{\theta}}
\rnc{\b}{\textbf}
\nc{\m}{\textrm}
\nc{\bb}{\mathbb}
\rnc{\bf}{\mathbf}
\nc{\til}{\texttildelow}
\nc{\be}{\begin{equation}}
\nc{\ee}{\end{equation}}
\nc{\dps}{\displaystyle}
\rnc{\l}{\left(}\rnc{\r}{\right)}
\nc{\lc}{\left\{}\nc{\rc}{\right\}}
\nc{\lb}{\left[}\nc{\rb}{\right]}
\nc{\ba}[1]{\begin{array}{#1}}
\nc{\ea}{\end{array}}       
\nc{\ra}{\rightarrow}
\nc{\Ra}{\Rightarrow}
\nc{\li}{\left |}
\nc{\ri}{\right |}
\nc{\pde}[2]{\frac{\partial #1}{\partial #2}}
\nc{\ode}[2]{\frac{d #1}{d #2}}
\nc{\odee}[3]{\frac{d^{#3} #1}{d #2^{#3}}}
\nc{\pdee}[3]{\frac{\partial^{#3} #1}{\partial #2^{#3}}}
\nc{\bn}{\begin{enumerate}}
\nc{\en}{\end{enumerate}}
\nc{\bt}{\begin{theorem}}
\nc{\et}{\end{theorem}}
\nc{\y}[1]{\lambda_{#1}}

\begin{document}
\mainmatter
\title{A graphical environment to express the semantics of computer-controlled systems}
\author{Timothy Wang 
\thanks{Timothy E. Wang is a PhD candidate in the Aerospace Engineering Department at Georgia 
Institute of Technology {\tt\small gtg176i@mail.gatech.edu}} 
\and Romain Jobredeux \thanks{Romain Jobredeaux is a PhD Student 
in the Aerospace Engineering Department at Georgia Institute of Technology 
{\tt\small rjobredeaux3@gatech.edu}} 
\and Eric Feron 
\thanks{Dr. Eric Feron is the Dutton-Ducoffe Professor of Aerospace Engineering at 
Georgia Institute of Technology \tt\small feron@gatech.edu}
\institute{Georgia Institute of Technology,\\
225 North Avenue, Atlanta, Georgia 30332-0001 }}

\maketitle
\begin{abstract}
We present the concept of a unified graphical environment for expressing the semantics of control systems. 
The graphical control system design environment in Simulink already allows engineers to insert a
variety of assertions aimed the verification and validation of the control software. 
We propose extensions to a Simulink-like environment's annotation capabilities to include 
formal control system stability, performance properties and their proofs. 
We provide a conceptual description of a tool, that takes in a Simulink-like diagram of the 
control system as the input, and generates a graphically annotated control system diagram as the output. 
The annotations can either be inserted by the user or generated automatically by a third party control 
analysis software such as IQC$\beta$ or $\mu$-tool.  
We finally describe how the graphical representation of the system and its properties can be translated to annotated programs in a 
programming language used in verification and validation such as Lustre or C. 
\end{abstract}

\section{Introduction}

\label{sec:introduction}
Embedded control systems are ubiquitous in present day safety-critical applications. The aerospace and 
medical fields are filled with examples of such systems. The verification and validation (V\&V) of 
their software implementation has always been a major preoccupation given the 
dire consequences of any potential malfunction. It has been the endeavor of the formal methods community
to provide tools that facilitate and rationalize this process. 
However, currently there is little communication between the engineers who design 
the control system and the engineers who do the V\&V. 
It has been noted in \cite{feron_csm} that the 
former could potentially provide valuable inputs for the latter in regards to 
finding the relevant invariants. 
We believe that inputs from the control engineer can be
helpful to the V\&V community if the following can be provided:
\begin{itemize}
\item An environment for the control engineers to easily insert stability and performance proofs
into their designs. 
\item Automatic translation of the information provided by the control engineers into a form 
that is familiar to the V\&V community. 
\end{itemize}
In this paper, we present an extension to the current block diagram representations of control 
systems (Simulink, Xcos) that include fundamental control systems proof 
information such as a Lyapunov function, which establishes stability, 
and the plant model with respect to which stability was established at the time the controller was 
designed. These extensions resemble, but are different from, Simulink's current diagram annotation 
capability.

\subsection{Challenges}

The following are the challenges that we like to address:
\begin{itemize}
\item Provide a coherent set of new blocks in a Simulink-like environment that enable a wide 
array of systems and types of stability proofs to be handled. 
\item Provide a formal semantics for the new graphical environment. The semantics can be inherited 
from a Simulink-like environment that has a formal semantics such as Scade. 
\item Develop a tool to perform translation from the Simulink-like environment to an industrial programming language such as Lustre or C. 
\end{itemize}

\subsection{Background}
Simulink being the de-facto industry tool for embedded controllers, 
has been the subject of many research efforts in the validation and verification community. 
There have been numerous attempts at the translation of Simulink into other languages for
the purpose of formal analysis. See \cite{translate00}, \cite{translate01},\cite{translate02} and
\cite{translate03}. 
The ideas presented in this paper are not specifically confined to Simulink, 
but rather they form an overall concept that can be implemented in any graphical modeling language tool. 
This work is closely related to the ideas presented in \cite{feron_csm}, as it explores autocoding 
with proof at the interface with the control engineer.

\section{Conceptual Overview}
\label{sec:overview}
\begin{figure}[htp]
\centering
\includegraphics[width=4.5in]{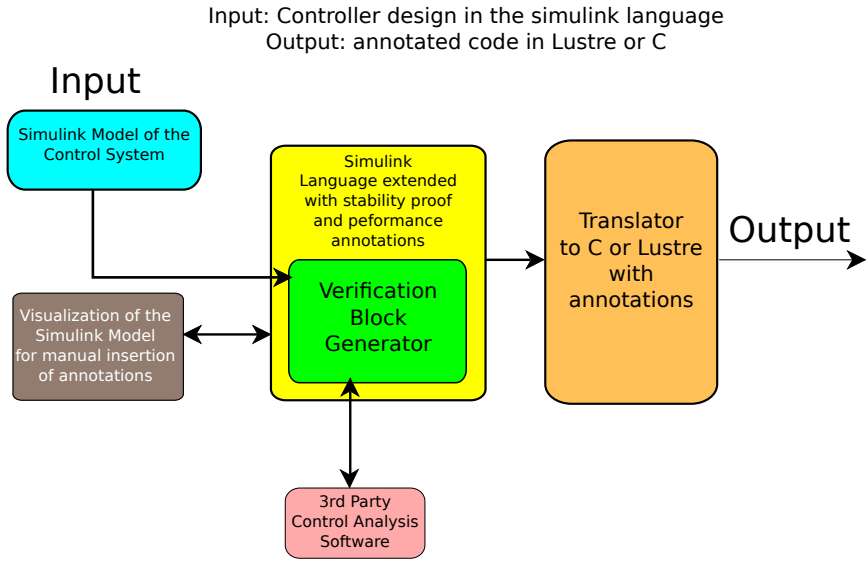}
\caption{Overall Picture}
\label{fig:tool_design}
\end{figure}

We begin by proposing a tool represented by the illustration in figure \ref{fig:tool_design}. 
The front-end of the tool provides a unified graphical environment for the design of control
systems as well as the insertion of proofs about the control systems. 
The back-end translates the visual design with proofs into annotated code in an industrial 
programming language, such as Lustre or C which can then be analyzed using 
V\&V tools developed by the formal methods community. 
We describe the top-level components that make up the tool with the focus on the graphical environment 
for the expression of control systems semantics. 
The demonstrations in this paper use existing annotative capabilities of the 
graphical modeling platform Simulink.
\subsection{Front-end}

\begin{figure}[htp]
\centering
\includegraphics[width=4.5in]{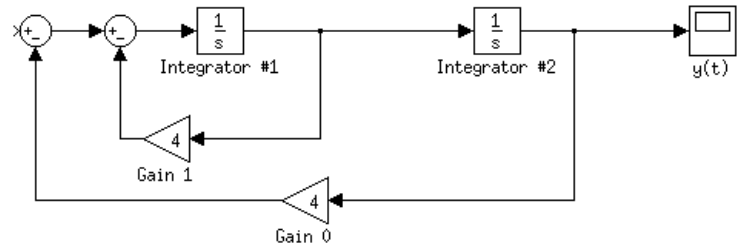}
\caption{Model Input}
\label{fig:simulink_model}
\end{figure}

\begin{figure}[htp]
\centering
\includegraphics[width=4.5in]{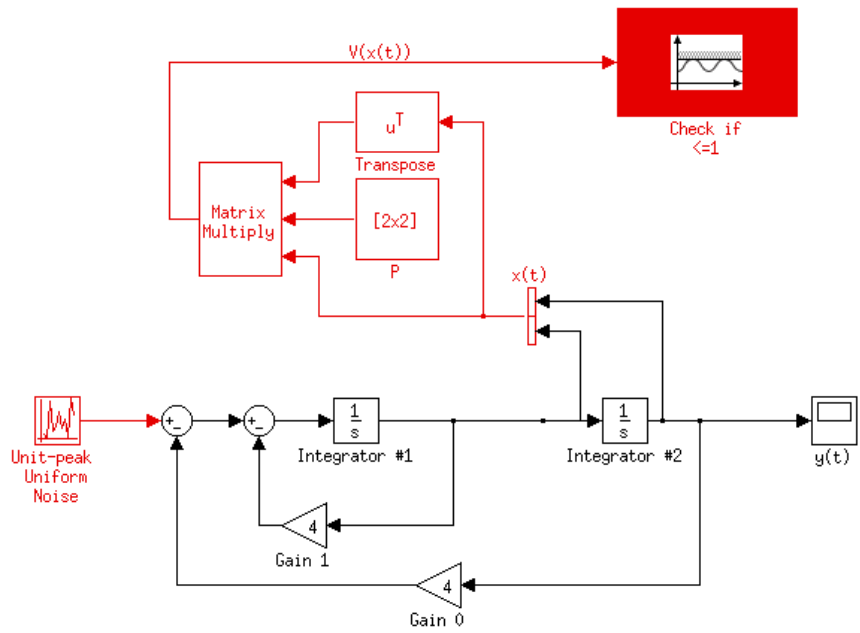}
\caption{Verification Block Generator Output}
\label{fig:bounded_noise}
\end{figure}

The front-end of the tool takes in an input of a Simulink-like model and 
feeds to the \emph{verification block generator} (the green block in figure \ref{fig:tool_design} ). 
Let us consider the model of the double integrator system in figure \ref{fig:simulink_model} as 
the input. 
The \emph{verification block generator} produces an output model that includes additional blocks 
and wires (see the red portion in figure \ref{fig:bounded_noise}), 
which are arranged to express an assertion on the states of the input model. 
The Simulink environment gives the user great flexibility in
expressing a variety of stability and performance criteria
from control theory literature. 
In this particular example the \emph{verification block}
checks the stability of the double integrator system in the presence of a noise that is bounded
in power. 
However, note that the stability proof annotation in figure \ref{fig:bounded_noise} 
was constructed by concatenating two signals
together, and then feeding the output
through several mathematical and logic blocks from the Simulink library. 
This can become a very cumbersome process to the control engineer. 
To simplify this as much as possible we propose an extension to the current Simulink block library to allow a more direct way of expressing control systems properties and their proofs. 
For example the bounded noise stability property in figure \ref{fig:bounded_noise} 
can be captured more succinctly by a new annotation block type denoted \emph{stability} with the following 
three parameters: the positive-definite matrix $P$, the characteristics of the noise input and the states of the control system. 
More examples and detailed descriptions of the variety of control systems properties and proofs can be found in section \ref{sec:examples}. 

\subsection{Automatic Generation of Control System Proofs}

The most important parameter in the proposed \emph{stability} block type is the positive-definite 
matrix $P$ since stability proofs for many control systems boils down to computing this matrix 
\cite{lmi_boyd}, and the proof flows down to the code-level in a nice fashion \cite{feron_csm}.   
To automate this process as much as possible we propose linking the 
\emph{verification block generator} with automatic control system analysis tools (see the pink block in figure \ref{fig:tool_design}). 
Several third party tools exist which can adapted for this role. 
For linear control systems the $P$ can always be generated automatically by 
a robust control toolbox such as the IQC$\beta$ or $\mu$-tool (see \cite{iqc_manual},\cite{iqc} and \cite{mutool}). 
For nonlinear control systems such as adaptive controllers 
it is likely that manual input will be needed.
For example the user might be expected to insert the Lyapunov function by using the 
appropriate Simulink blocks and connecting them with the wires. 
For these cases where the automation fails, the model output from the \emph{verification 
block generator} is provided 
to the user as the interface for the manual insertions of the stability proofs (see
the gray block in figure \ref{fig:tool_design}). 

\subsection{Formalism}
We propose a new type of wire that elevates signals to the abstract level to allow
easy differentiation between signals that represent the states of the control system i.e. $x(t)$ in 
figure \ref{fig:bounded_noise} and signals
that are not the states of the control system.  
The rest of the newly proposed annotation block types can use the existing blocks and wires structure, 
with some indication that they are not to be used for code generation, but for annotation generation. 
For this we also propose an intermediate language representation of the Simulink model that makes this
distinction clearly. 
The existing labeling options that exist in Simulink can be useful here to keep track of names.

\subsection{Back-end}
The back-end of the tool is the translator from the graphical environment to a programming
language such as C or Lustre.  
The majority of work here is will be formalizing the semantics of the graphical environment. 
The translation process must also deal with time discretization of the stability proof annotations, which 
can be quite difficult depending on the type of control system that the user put into the tool. 

\section{Simulink Examples}
\label{sec:examples}
We present several examples of annotating stability proofs and performance criteria for
control systems in the Simulink environment. These examples can obtained from control engineering
books such as \cite{naira} or \cite{haddad}. 

\subsection{Semantics of the Simulink blocks and wires}

The mathematical operational blocks such as sum, multiply, divide, dot product
are polymorphic just like every other Simulink blocks.
They take input arguments of many different types: scalar, vector of arbitrary dimensions, and matrices. 
Semantically the blocks can change either due to different input arguments or user specification. 
For example the "Product" block can be a product of scalars, matrix multiplication, or element-wise multiplication of 
the entries of the vector depending
on the input argument and user choice. 
The wires can carry all numerical data types available in Simulink.  The two relevant types 
in the control system diagrams are scalars and vectors that are assumed to be either real or complex. 
For expressing most annotations of stability proofs and performance criteria,
we can use the existing blocks in Simulink. 
A small amount of annotations may require the convenience of 
functions defined outside of Simulink environment i.e. MATLAB for example. 

\subsection{Lyapunov Stability}

\begin{figure}[t]
\centering
\includegraphics[width=4.2in]{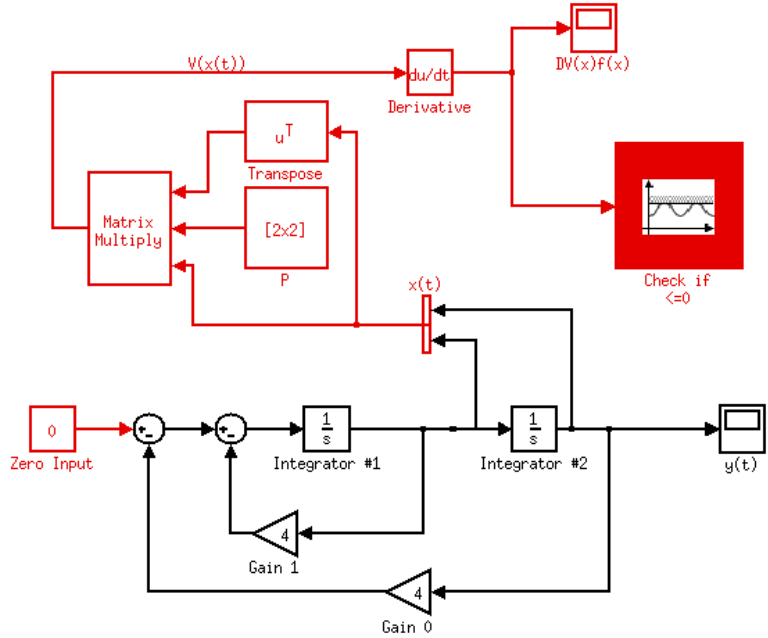}
\caption{Lyapunov Stability}
\label{fig:lyapunov}
\end{figure}

We start with Lyapunov stability, since it is the simplest stability result in 
time-domain that we can express using the graphical environment of Simulink. 
The essential part of the Lyapunov stability is the quadratic 
Lyapunov function $V(x)$ in (\ref{lyap01}) where $x$ is the 
state of the control system and $P$ is a positive-definite matrix. 
\be
\dps V(x) =  x^{\m{T}} P x 
\label{lyap01}
\ee
The noise block is set to $0$. 
The \emph{verification block} shown in figure \ref{fig:lyapunov} expresses the following assertion: 
the lie derivative of the Lyapunov function $V(x(t))$ is less than or equal to zero. 
The matrix $P$ can be computed using either one of the two robust control toolboxes mentioned. 

\subsection{$\mathcal{L}_{2}$ Gain Stability}

\begin{figure}[t]
\centering
\includegraphics[width=4.2in]{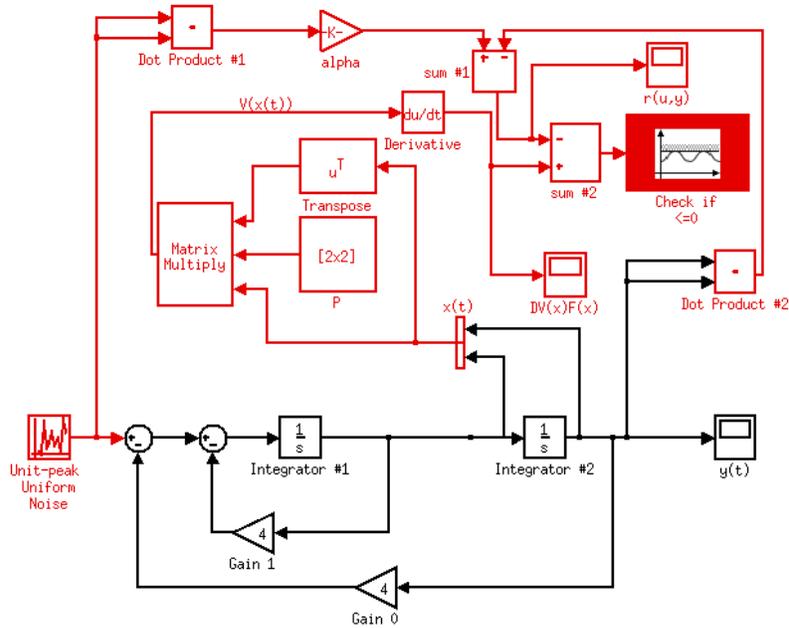}
\caption{Finite Gain $\mathcal{L}_{2}$ Stable}
\label{fig:l2gain}
\end{figure}

For input-to-output stability, the annotation diagram becomes more complex.  In this case 
both the storage and supply functions need to be constructed \cite{haddad}. Figure \ref{fig:l2gain} has
an example of a proof annotation showing finite $\mathcal{L}_{2}$-gain stability of the double integrator
system. 
The annotation expresses the following assertion:
\be
\dps \dot{V}(x) - \alpha^2 w^{\m{T}} w + y^{\m{T}} y \leq 0
\label{l2gain_ineq}
\ee
where $w$ is the unit-peak uniform noise input, $V(x)=x^{\m{T}} P x$ is a quadratic Lyapunov-like function called
the storage function, and $y$ is the output signal. 

The proposed extensions to the interface will reduce drastically the number of blocks and wires 
necessary to construct this stability proof annotation. 
Despite the increase in complexity one of the essential component of the annotation is still a positive-definite matrix $P$. 
This $P$ can also be obtained using the 
two robust control solvers mentioned in section \ref{sec:overview}, therefore the diagram in figure \ref{fig:l2gain} 
can be generated automatically. 

\subsection{Plant Model as Annotation}
\begin{figure*}[htp]
\centering
\includegraphics[width=5.2in]{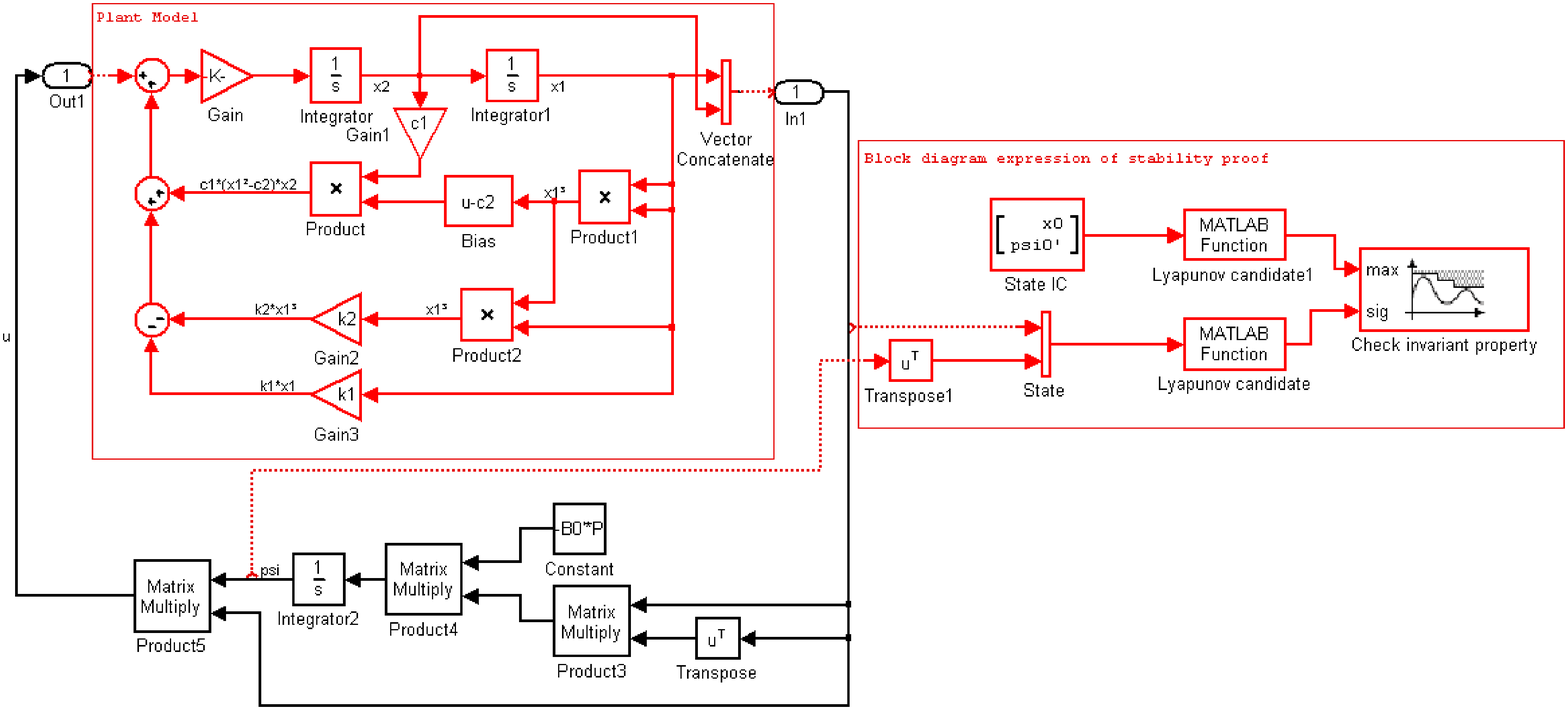}
\caption{Adaptive controller with Plant}
\label{fig:plant}
\end{figure*}

Most controller stability proofs are based on some form of model for the plant that is being controlled. 
Thus the plant needs to be introduced somehow in the annotation framework. 
This is actually relatively straightforward at the graphical level, 
since most graphical modeling tools are not only meant for implementing controllers, but also for testing them, 
and in the latter case a model for the plant must be present. Figure \ref{fig:plant} shows how we go about displaying plant model information at an abstract level, which will not interfere with the executable code that will be generated.
We take the example of the following plant from \cite{hayakawa}
\be
\dps m \ddot{x} + c_1 \l x^2-c_2 \r\dot{x} + \l k_1+k_2 x^2 \r x = u
\ee
with the adaptive controller
\be\ba{c}
\dps u=\psi x \cr
\dps \dot{\psi}=-B_{0}^{\m{T}} P \lb \ba{cc} x^2  & x\dot{x} \cr x\dot{x} & \dot{x}^2 \ea \rb 
\ea
\label{adaptive_law}
\ee
The Lyapunov stability of the feedback interconnection of the plant and the controller 
is established by the Lyapunov function
\be\ba{lcl}
\dps V(x,\dot x, \psi)& = &\frac{1}{2} \lb 
\ba{c} x \cr \dot{x} \ea \rb^{\m{T}} P 
\lb \ba{c} x \cr \dot{x} \ea \rb  +  p_{2} \dps \int_0^x  \sigma \tilde{K}(\sigma) d\sigma \cr
& + & \dps p_{12} \int_0^x  \sigma \tilde{C}(\sigma) 
d\sigma + \frac{1}{2} \psi \psi^{\m{T}}
\ea
\label{lyap03}
\ee

This complex example is mainly introduced to show the almost unlimited expressive power of the proposed interface extension.  
\subsection{$\mathcal{L}_{1}$ Adaptive Controller Performance} 

\begin{figure}[htp]
\centering
\includegraphics[width=4.5in]{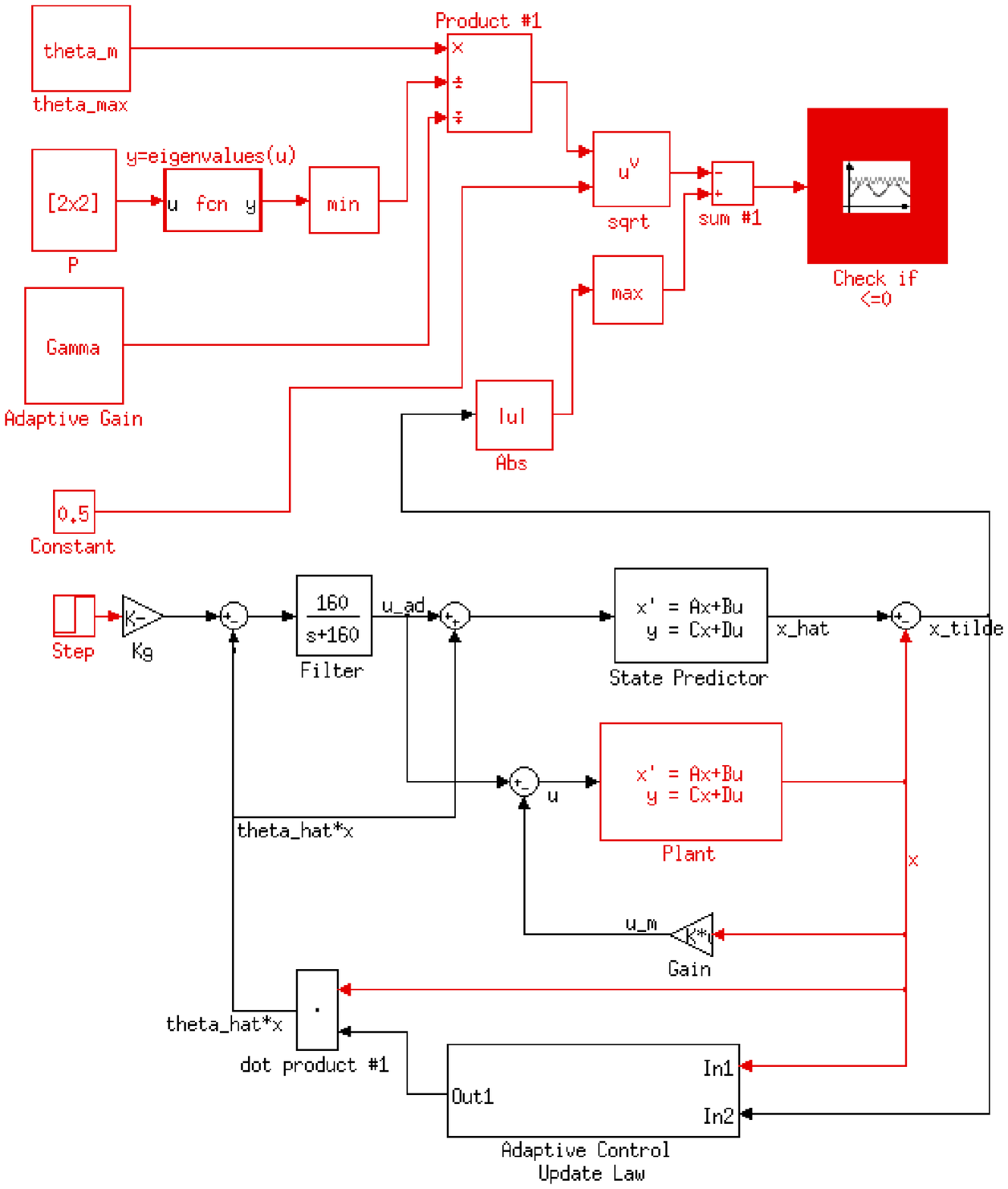}
\caption{$\mathcal{L}_{1}$ Transient Performance Bound}
\label{fig:l1_model}
\end{figure}

To further demonstrate the expressive power of the graphical environment, we show
the following $\mathcal{L}_{1}$ adaptive controllers (see figure \ref{fig:l1_model}) and an example of annotating a
performance bound.  For $\mathcal{L}_{1}$ controllers there exists a proof of not only closed-loop 
Lyapunov stability but also bounds on the transient performances.  For example the bound on the state
prediction error $\tilde{x}$ of the controller is a function of the uncertainty of the plant "theta\_max",
the Lyapunov function matrix $P$ and the adaptive gain "Gamma". Note that the uncertainty parameter of plant can be part of the plant model or separate by itself as in figure \ref{fig:l1_model}. 
\be
\dps \| \tilde{x} \|_{\mathcal{L}_{\infty}}-\sqrt{\frac{\theta_{max}}{\lambda_{min}\l P \r \Gamma}} \leq 0
\label{l1_bound}
\ee

\subsection{Discretization during Translation}

Thus far we have shown several examples of continuous-time proofs of stability for 
control systems. Now how do these proofs translate down to the code-level?
The main issue as mentioned before is the time-discretization that is applied to the model when it 
is translated into code. 
Fortunately for all linear control systems, the discrete-time stability proofs can be easily and systematically obtained 
However for the adaptive controllers, it is very impractical to look for
Lyapunov stability proofs in discrete-time hence the problem remains that we must use the continuous-time proofs as invariants for the discretized system.

\section{Integration with Third Party Tools}
\label{sec:integration}

For the integration of third party control system analysis tools, we'll need 
to build a component that extracts the necessary control system parameters and characteristics
from the input Simulink model. For this procedure to be automated 
it is again necessary to have a formal semantics of the 
proposed graphical environment. 
System information such as the state-space model, the system states, the input noise 
disturbance are examples of the required inputs
needed for IQC$\beta$ and $\mu$-tool. 

\section{Conclusion and Future Work}

In this paper we have proposed a new graphical environment that allows the easy
expression of the semantics of computer-controlled systems. We believe this environment
simplifies the process by which the control engineer can provide domain knowledge
for the deductive verification of the controller implementation.  
We have provided several examples of how control stability proofs and performance
criteria can be expressed in a current graphical modeling environment i.e. Simulink. 
For the new graphical environment, we have proposed several extensions designed 
to enhance the proof annotative capabilities of the current environment. 
We are currently in the process of formalizing the new unified graphical environment.  
This is a proposed research orientation that 
still requires much effort: formalizing the graphical annotation "language", 
integrating it in an existing graphical modeling environment, 
interfacing it with third party tools, and lastly
implementing the translation tool that will generate 
the annotated code from the extended diagrams. 
\section{Acknowledgments}

This research is supported by the Northeastern University and NASA,
by the United States Army through the contract W911NF-11-10046, and
by Aurora Flight Sciences through the contract AFS10-0036. 

\bibliographystyle{abbrv}
\bibliography{fm}

\end{document}